%% file: main.tex
\documentclass[conference,letterpaper]{IEEEtran}

\usepackage[utf8]{inputenc} 
\usepackage[T1]{fontenc}
\usepackage{url}
\usepackage{ifthen}
\usepackage{cite}
\usepackage[cmex10]{amsmath} 
\usepackage{xcolor}
\usepackage{amssymb}
\usepackage{graphicx}
\usepackage[font=scriptsize]{caption}

\usepackage{graphicx}
\usepackage{hyperref} 
\usepackage{amsmath} 
\usepackage{xfrac}
\usepackage{amssymb}
\usepackage{mathtools}
\usepackage{tikz}
\usetikzlibrary{positioning}
\usetikzlibrary{trees}
\usetikzlibrary{shapes,arrows}
\usetikzlibrary{arrows}
\usetikzlibrary{arrows,positioning, calc}

\newcommand{\figref}[1]{Fig.~\ref{#1}}  

\begin{document}
\title{Nested Construction of Polar Codes via Transformers} 


\author{%
  \IEEEauthorblockN{Sravan Kumar Ankireddy}
  \IEEEauthorblockA{University of Texas at Austin}\\
 
  \and

  \IEEEauthorblockN{S. Ashwin Hebbar}
  \IEEEauthorblockA{Princeton University}\\
 
\and
  
  \IEEEauthorblockN{Heping Wan, Joonyoung Cho, and Charlie Zhang}
  \IEEEauthorblockA{Samsung Research America}
}


\maketitle


\let\emptyset\varnothing

\input{abstract}

\input{intro}

\input{problem}

\input{algo}

\input{related_works}

\input{results}

\input{conclusion}

\clearpage
\input{bibilography}

\clearpage
\normalsize
\input{appendix}


\end{document}

%% file: abstract.tex
\begin{abstract}
     Tailoring polar code construction for decoding algorithms beyond successive cancellation has remained a topic of significant interest in the field. However, despite the inherent nested structure of polar codes, the use of sequence models in polar code construction is understudied. In this work, we propose using a sequence modeling framework to iteratively construct a polar code for any given length and rate under various channel conditions. Simulations show that polar codes designed via sequential modeling using transformers outperform both 5G-NR sequence and Density Evolution based approaches for both AWGN and Rayleigh fading channels.
    
\end{abstract}

%% file: intro.tex
\section{Introduction}\label{sec:intro}

Polar codes are a significant breakthrough in 21st-century information theory, being the first deterministic code construction to achieve capacity for binary memoryless symmetric channels \cite{arikan2009channel}. Besides, they also achieve excellent reliability in the short-to-medium blocklength regime with a relatively low-complexity encoding scheme. Their practical impact is underscored by their adoption in 5G standards, in control channels, with potential application in data channels in future generations \cite{dahlman20205g}.

While polar codes achieve capacity under successive cancellation (SC) decoding, many improvements were made subsequently to improve the performance at short block lengths, most notably successive cancellation list (SCL) decoding \cite{tal2015list, balatsoukas2015llr}. In practical applications, a version of the Polar code concatenated with a cyclic redundancy check (CRC) is commonly used, paired with CRC-aided successive-cancellation list (CA-SCL) decoding to enhance performance. The underlying principle of Polar codes is channel polarization, realized through the recursive application of the Polar kernel. This process creates sub-channels under successive cancellation decoding with either very high or very low capacity. An efficient coding strategy involves using high-capacity sub-channels for information transmission while freezing the rest. 

More recently, the finite blocklength error correction capability of Polar codes has been further improved by pre-transformed polar coding schemes like Polarization-adjusted convolutional (PAC) codes \cite{arikan2019sequential, trifonov2015polar, li2019pre, choi2023deep}. Notably, PAC codes approach the finite-blocklength channel capacity by allowing for more effective utilization of bit-channel capacities. However, the requirement of high complexity decoders has impeded practical implementation. 



The construction of a polar code for any channel involves ranking the reliability of the polarized or transformed sub-channels. Once the ranking is performed, polar encoding involves selecting the top $K$ reliable positions out of $N$ available positions to construct a rate $\sfrac{K}{N}$ code. Historically, methods such as Monte-Carlo \cite{arikan2009channel}, Density Evolution (DE) \cite{mori2009performance, tal2013construct}, DE with Gaussian Approximation (GA) \cite{trifonov2012efficient} have been proposed to generate this reliability sequence for AWGN channels \emph{under SC decoding}. Subsequent efforts have also been made to approximate channel reliabilities under the SCL decoder ~\cite{mondelli2014polar,qin2017polar,miloslavskaya2020design}. However, these approaches are constrained by their dependence on specific rates and channel conditions, necessitating recalibration of the algorithms for different scenarios. Addressing this challenge, the 3GPP standards \cite{3gpp2018multiplexing} have introduced universal reliability sequence that performs satisfactorily across various rates, blocklengths, and channel conditions \cite{3gpp2018multiplexing, bioglio2020design} while maintaining a low complexity, termed a \textit{nested polar code}. Despite its versatility, this sequence is known to be sub-optimal for many shorter block lengths and is less studied for channels other than AWGN.

In this work, we propose a novel approach that utilizes a sequential decision-making algorithm for a nested polar code construction under a CA-SCL decoder, applicable to different channels such as AWGN and fading channels. Our focus is on examining the potential of recent advancements in sequence modeling and deep learning, particularly the application of self-attention-based transformer architecture to model the underlying reliability sequence in a purely data-driven fashion. 
Our key contributions can be summarized as follows:
\begin{itemize}
    \item We propose a sequential decision-making algorithm based on a transformer to construct nested polar codes.
    \item We study the effectiveness of a learning-based design for non-AWGN channels, demonstrating higher gains over the 5G-NR sequence, particularly on fading channels.
    \item We highlight the importance of preserving the positions of indices in input representations, as opposed to using permutation-invariant representations, via a systematic ablation study. 
\end{itemize} 


%% file: problem.tex
\section{Problem Formulation and Background}\label{sec:problem}
\subsection{Polar codes}

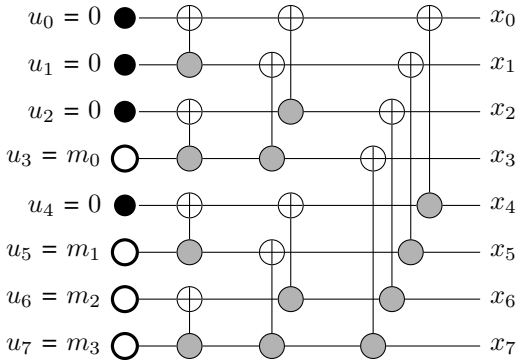
\begin{figure}
\centering
\begin{tikzpicture}[
rect/.style={rectangle, draw=black!100, fill=white!5, thick, minimum width=0.8cm, minimum height=0.5cm},
circ_info/.style={circle, draw=black!100, fill=black!0, very thick, minimum size=0mm},
circ_frozen/.style={circle, draw=white!100, fill=black!100, very thick, minimum size=0mm},
circ/.style={draw,circle, fill=black!30, minimum size=0mm},
empty/.style={rectangle, draw=white!100, fill=white!5, thick, minimum width=0cm, minimum height=0cm},
xor/.style={draw,circle,  minimum size=0mm, append after command={
        [shorten >=\pgflinewidth, shorten <=\pgflinewidth,]
        (\tikzlastnode.north) edge (\tikzlastnode.south)
        (\tikzlastnode.east) edge (\tikzlastnode.west)
        }
    }
]
\node[circ_frozen] (ch0_0_in) [label=left:{$u_0$ = $0$}] {};
\node[circ_frozen] (ch0_1_in) [below = 0.25cm of ch0_0_in] [label=left:{$u_1$ = $0$}] {};
\node[circ_frozen] (ch0_2_in) [below = 0.25cm of ch0_1_in] [label=left:{$u_2$ = $0$}] {};
\node[circ_info] (ch0_3_in) [below = 0.25cm of ch0_2_in] [label=left:{$u_3$ = $m_0$}] {};
\node[circ_frozen] (ch0_4_in) [below = 0.25cm of ch0_3_in] [label=left:{$u_4$ = $0$}] {};
\node[circ_info] (ch0_5_in) [below = 0.25cm of ch0_4_in] [label=left:{$u_5$ = $m_1$}] {};
\node[circ_info] (ch0_6_in) [below = 0.25cm of ch0_5_in] [label=left:{$u_6$ = $m_2$}] {};
\node[circ_info] (ch0_7_in) [below = 0.25cm of ch0_6_in] [label=left:{$u_7$ = $m_3$}] {};

\node[xor] (ch1_0_in) [right = 0.5cm of ch0_0_in] {};
\node[circ] (ch1_1_in) [right = 0.5cm of ch0_1_in] {};
\node[xor] (ch1_2_in) [right = 0.5cm of ch0_2_in] {};
\node[circ] (ch1_3_in) [right = 0.5cm of ch0_3_in] {};
\node[xor] (ch1_4_in) [right = 0.5cm of ch0_4_in] {};
\node[circ] (ch1_5_in) [right = 0.5cm of ch0_5_in] {};
\node[xor] (ch1_6_in) [right = 0.5cm of ch0_6_in] {};
\node[circ] (ch1_7_in) [right = 0.5cm of ch0_7_in] {};

\node[xor] (ch2_0_in) [right = 1cm of ch1_0_in] {};
\node[xor] (ch2_1_in) [right = 0.75cm of ch1_1_in] {};
\node[circ] (ch2_2_in) [right = 1cm of ch1_2_in] {};
\node[circ] (ch2_3_in) [right = 0.75cm of ch1_3_in] {};
\node[xor] (ch2_4_in) [right = 1cm of ch1_4_in] {};
\node[xor] (ch2_5_in) [right = 0.75cm of ch1_5_in] {};
\node[circ] (ch2_6_in) [right = 1cm of ch1_6_in] {};
\node[circ] (ch2_7_in) [right = 0.75cm of ch1_7_in] {};

\node[xor] (ch3_0_in) [right = 1.5cm of ch2_0_in] {};
\node[xor] (ch3_1_in) [right = 1.5cm of ch2_1_in] {};
\node[xor] (ch3_2_in) [right = 1cm of ch2_2_in] {};
\node[xor] (ch3_3_in) [right = 1cm of ch2_3_in] {};
\node[circ] (ch3_4_in) [right = 1.5cm of ch2_4_in] {};
\node[circ] (ch3_5_in) [right = 1.5cm of ch2_5_in] {};
\node[circ] (ch3_6_in) [right = 1cm of ch2_6_in] {};
\node[circ] (ch3_7_in) [right = 1cm of ch2_7_in] {};

\node[empty] (ch4_0_in) [right = 0.5cm of ch3_0_in] {$x_0$};
\node[empty] (ch4_1_in) [right = 0.75cm of ch3_1_in] {$x_1$};
\node[empty] (ch4_2_in) [right = 1cm of ch3_2_in] {$x_2$};
\node[empty] (ch4_3_in) [right = 1.25cm of ch3_3_in] {$x_3$};
\node[empty] (ch4_4_in) [right = 0.5cm of ch3_4_in] {$x_4$};
\node[empty] (ch4_5_in) [right = 0.75cm of ch3_5_in] {$x_5$};
\node[empty] (ch4_6_in) [right = 1cm of ch3_6_in] {$x_6$};
\node[empty] (ch4_7_in) [right = 1.25cm of ch3_7_in] {$x_7$};

\draw[-] (ch0_0_in.east) -- (ch1_0_in.west);
\draw[-] (ch0_1_in.east) -- (ch1_1_in.west);
\draw[-] (ch0_2_in.east) -- (ch1_2_in.west);
\draw[-] (ch0_3_in.east) -- (ch1_3_in.west);
\draw[-] (ch0_4_in.east) -- (ch1_4_in.west);
\draw[-] (ch0_5_in.east) -- (ch1_5_in.west);
\draw[-] (ch0_6_in.east) -- (ch1_6_in.west);
\draw[-] (ch0_7_in.east) -- (ch1_7_in.west);

\draw[-] (ch1_0_in.east) -- (ch2_0_in.west);
\draw[-] (ch1_1_in.east) -- (ch2_1_in.west);
\draw[-] (ch1_2_in.east) -- (ch2_2_in.west);
\draw[-] (ch1_3_in.east) -- (ch2_3_in.west);
\draw[-] (ch1_4_in.east) -- (ch2_4_in.west);
\draw[-] (ch1_5_in.east) -- (ch2_5_in.west);
\draw[-] (ch1_6_in.east) -- (ch2_6_in.west);
\draw[-] (ch1_7_in.east) -- (ch2_7_in.west);

\draw[-] (ch2_0_in.east) -- (ch3_0_in.west);
\draw[-] (ch2_1_in.east) -- (ch3_1_in.west);
\draw[-] (ch2_2_in.east) -- (ch3_2_in.west);
\draw[-] (ch2_3_in.east) -- (ch3_3_in.west);
\draw[-] (ch2_4_in.east) -- (ch3_4_in.west);
\draw[-] (ch2_5_in.east) -- (ch3_5_in.west);
\draw[-] (ch2_6_in.east) -- (ch3_6_in.west);
\draw[-] (ch2_7_in.east) -- (ch3_7_in.west);

\draw[-] (ch3_0_in.east) -- (ch4_0_in.west);
\draw[-] (ch3_1_in.east) -- (ch4_1_in.west);
\draw[-] (ch3_2_in.east) -- (ch4_2_in.west);
\draw[-] (ch3_3_in.east) -- (ch4_3_in.west);
\draw[-] (ch3_4_in.east) -- (ch4_4_in.west);
\draw[-] (ch3_5_in.east) -- (ch4_5_in.west);
\draw[-] (ch3_6_in.east) -- (ch4_6_in.west);
\draw[-] (ch3_7_in.east) -- (ch4_7_in.west);

\draw[-] (ch1_0_in.south) -- (ch1_1_in.north);
\draw[-] (ch1_2_in.south) -- (ch1_3_in.north);
\draw[-] (ch1_4_in.south) -- (ch1_5_in.north);
\draw[-] (ch1_6_in.south) -- (ch1_7_in.north);

\draw[-] (ch2_0_in.south) -- (ch2_2_in.north);
\draw[-] (ch2_1_in.south) -- (ch2_3_in.north);
\draw[-] (ch2_4_in.south) -- (ch2_6_in.north);
\draw[-] (ch2_5_in.south) -- (ch2_7_in.north);

\draw[-] (ch3_0_in.south) -- (ch3_4_in.north);
\draw[-] (ch3_1_in.south) -- (ch3_5_in.north);
\draw[-] (ch3_2_in.south) -- (ch3_6_in.north);
\draw[-] (ch3_3_in.south) -- (ch3_7_in.north);
\end{tikzpicture}
\caption{Encoding of (8,4) Polar code with frozen nodes set to "$0$"}
\label{fig:encoding}

\end{figure}

Polar codes are a class of codes characterized by the phenomenon of channel polarization. This process transforms $N$ copies of a binary discrete memoryless channel (B-DMC), denoted as $W(Y|X)$, into $N$ synthetic "bit channels." These bit channels are \textit{polarized} - i.e., each of them can transmit a single bit with different reliabilities. The bit channels diverge in their reliability for large enough N, becoming either completely noisy or completely noiseless. 
The basic building block of Polar codes is the polarization kernel \cite{korada2010polar}, which can be represented by the matrix $G_2 = \begin{bmatrix} 1 & 0 \\ 1 & 1 \end{bmatrix}$, transforming $(u, v)$ into $(u \oplus v, v)$, where $\oplus$ denotes the XOR operation. The generative matrix $G_N$, is constructed as the $n$-fold Kronecker product of $G_2$ : $G_N = G_2^{\otimes n}$. To encode message bits $\textbf{m} = m_0, m_1, \cdots, m_{K-1}$, it is embedded into a source vector u, such that $u_\mathcal{F} = 0, u_\mathcal{F^C} = m$. The polar transform is efficiently realized using a tree-based procedure described in \figref{fig:encoding}. Polar codes can be specified completely by $(N, K, \mathcal{F})$, where $N$ is the block length, $K$ is the message length, and $\mathcal{F}$ is the set of frozen message positions; $|\mathcal{F}| = N-K$. By strategically selecting a subset of the most reliable bit channels for transmission of information bits and "freezing" the less reliable channels with known values (typically zeros), Polar codes effectively leverage channel polarization to maximize data throughput while minimizing the probability of decoding errors. 

The decoding of Polar codes is typically performed using successive cancellation (SC) decoding or its variants. The SC decoder operates in a sequential manner, estimating each bit of the transmitted message one at a time. Concretely, for each bit $u_i, i \notin \mathcal{F}$, the decoder computes the conditional probability $P(u_i | y, \hat{u_0}, \cdots \hat{u}_{i-1})$ to estimate $u_i$. An extension of this method to a breadth-first search, list decoding, is used in practice for enhanced decoding performance.

\subsection{Polar Code Construction}

The problem of constructing a polar code can be explicitly stated as finding the optimal set of frozen positions, $\mathcal{F}$, to minimize the block error rate for a given B-DMC $W(Y|X)$ and a decoding algorithm $\mathcal{D}$. This task involves selecting $K$ out of $N$ reliable bit channel channels. Traditional methods estimate and rank the reliability of each bit channel, choosing the most reliable ones. However, these approaches, such as Monte Carlo simulations \cite{arikan2009channel} and density evolution, are computationally expensive. Further, unlike most channel codes whose definition is independent of the channel and signal-to-noise ratio (SNR), the optimal $\mathcal{F}$ for Polar codes is contingent on both the channel characteristics and the choice of decoder. This dependency leads to high computational costs and hinders real-time computation of $\mathcal{F}$.

Deep learning (DL) has revolutionized progress in many fields, such as text and image processing. More recently, deep learning has shown immense potential to solve intractable problems in wireless communications, including non-linear channel code design~\cite{kim2018communication,kim2018deepcode,makkuva2021ko,jamali2022productae}, neural channel decoding~\cite{nachmani2016learning,nachmani2018deep,shlezinger2020viterbinet,choukroun2022error,hebbar2022tinyturbo,hebbar2023crisp, 
aharoni2023data, ankireddy2023interpreting}, and channel estimation~\cite{wen2018deep,balevi2020massive}. Indeed, the problem of designing a $(N,K)$ polar code can be viewed through the lens of DL as designing a model, say a neural network parameterized as $\Phi$, that inputs code parameters and channel conditions to identify the optimal $\mathcal{F}$ that minimizes the block error rate (BLER). 
\begin{align}
    \mathcal{F}_\Phi^* = \underset{\Phi}{\mathrm{argmin}} \; \text{BLER}(\Phi; N, K, W, \mathcal{D})
\end{align}

One approach is to view polar code construction as a combinatorial optimization problem, seeking the selection of the optimal $K$ out of $N$ positions. Techniques such as the Gumbel-Softmax trick \cite{jang2016categorical, herrmann2020channel} enable gradient propagation to discrete bit indices. However, this method faces scalability issues and involves backpropagation of gradients through the decoder, which is often infeasible.
As an alternative, reinforcement learning (RL) has been employed to model the construction algorithm's policy more efficiently, using reward signals instead of backpropagating gradients through the decoder. A comprehensive comparison of our RL-based method with existing approaches is detailed in Section \ref{sec:related}.

Notably, these conventional approaches require separate construction algorithm runs for each rate and SNR. This challenge was addressed by the 5G standardization, which introduced a unique universal bit-channel reliability sequence (1024 bit-channel indices sorted in reliability order) that can be used to design $\mathcal{F}$ for any polar code of length $<1024$ \cite{bioglio2020design}. This sequence, sorted by reliability, performs satisfactorily across various SNRs and rates. Such an approach, known as a \textit{nested polar code}, provides flexibility and scalability. Our work adopts a similar sequential modeling approach for constructing the information set, resulting in an efficient nested polar code structure, further discussed in the subsequent sections.



\subsection{Nested Construction using sequence modeling}

A desired property for polar code design is its nested nature, which ensures that a rate $sfrac{i}{N}$ code can be directly constructed from a rate $\sfrac{K}{N}$ by simply selecting the best $i$ indices among the $K$ indices. It is non-trivial to incorporate this property when viewing this as a combinatorial problem. Moreover, a brute force search over the combinatorial space is infeasible, as the cardinality of the space becomes intractable, growing as $N!$ with respect to blocklength $N$ growing to a size of $10^{88}$ even for N=64.

In line with the convention established in ~\cite{li2021learning}, let us consider $\{\pi_i \}_{j=1}^{K}$ as the set of $K$ information positions, resulting in a $\sfrac{K}{N}$ code. The loss (negative reward) for this rate, the BLER, is denoted by $J_k$. To integrate the nested structure into our modeling, we iteratively calculate the BLER for all rates $\{ \sfrac{1}{N}, \sfrac{2}{N}, \dots \sfrac{K}{N} \}$. Each code for a specific rate is constructed by considering the first $i$ indices from the nested code. We also introduce a weighting factor $c_k$, enabling a balance in performance between different rates. The nested code of rate $\sfrac{K}{N}$ can thus be constructed by the following optimization,
\begin{align}
    \min_{\pi} \sum_{i=1}^{K} c_k J_k (\{ \pi_j \}_{j=1}^{K}).
\end{align}
This transforms the problem of combinatorial optimization into a sequence modeling task of iteratively estimating one information position at a time. In this work, we aim to construct nested codes of rate from $1/N$ to $1$ unless specified otherwise, \textit{i.e,} $K=N$.

\begin{figure*}[!htb]
    \centering
 	\includegraphics[width=1.0\linewidth]{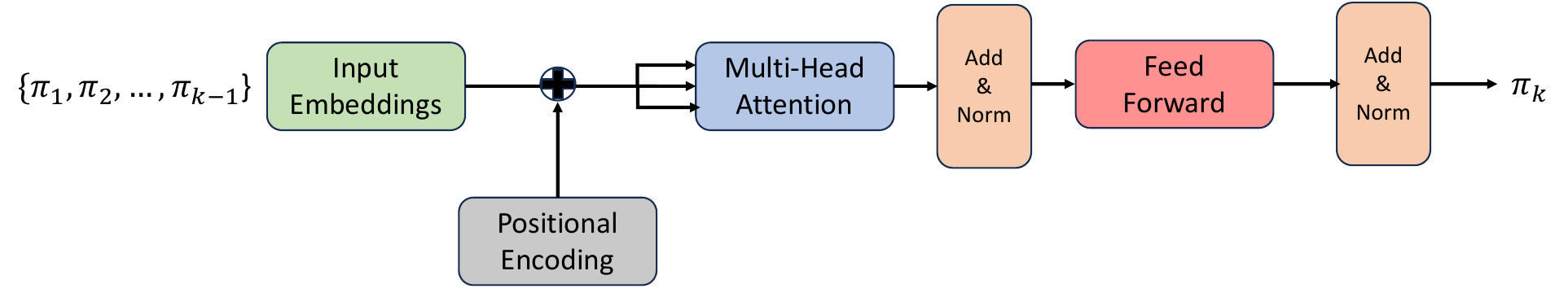}
 	\captionsetup{font=small}
 	\caption{Predicting the $k^{\text{th}}$ information position using the first $k-1$ predictions via transformer model.}
 	\label{fig:sysmodel}
\end{figure*}

%% file: algo.tex
\section{Sequential modeling for Polar Codes}


Reinforcement learning (RL) is a paradigm where an agent learns to make decisions by interacting with an environment. The process is formalized as a Markov Decision Process (MDP), defined by a tuple $(S, A, P, R)$, where $S$ is a set of states, $A$ is a set of actions, $P$ is the state transition probability matrix (i.e., $P(s_{t+1}|s_t, a_t)$ represents the probability of transitioning to state $s_{t+1}$ from state $s_t$ after taking action $a_t$), $R$ is the reward function (i.e., $R(s_t, a_t)$ is the reward received after taking action $a_t$ in state $s_t$). In our problem, the state is defined by the set of previously selected indices for polar code, the action space comprises the set of remaining indices to be selected, and the reward at each step is the negative BLER of the code formed by the selected indices. We model this as a deterministic environment with defined state transitions, employing a policy gradient approach to optimize the nested code construction.

The goal in RL is to learn a policy $\pi$, which is a mapping from states to actions, that maximizes the expected cumulative reward, formally defined as the expected return $G_t = \sum_{k=0}^{\infty} \gamma^k R_{t+k+1}$, where $R_{t+k+1}$ is the reward at time step $t+k+1$. Policy Gradient methods are a class of algorithms in RL that optimize the policy directly. In our approach, a policy network parameterized by $\Phi$ directly estimates the policy $\pi(a|s, \Phi)$, which outputs the next bit to be selected for the polar code sequence. This network is trained to maximize the expected cumulative reward. The loss function for the policy network is given by (Policy Gradient Theorem \cite{sutton2018reinforcement}):

\begin{equation}
L(\Phi) = -\mathbb{E}_{\pi_\Phi}[\log \pi(a|s, \Phi) G_t]
\end{equation}\label{eqn:policy_loss}

In this work, we model the policy network as a transformer encoder-only model \cite{vaswani2017attention, devlin2018bert}, as represented in Figure \ref{fig:sysmodel}. We view the objective as a  sequence modeling problem, similar to next-word prediction in language models. Starting with a null set $\emptyset$, we iteratively predict the best information position available among the remaining positions, one at a time. The positional encoding produces representations that preserve the order of previously predicted indices. We find the best policy for this objective by minimizing the policy gradient loss in \ref{eqn:policy_loss} to model the conditional probability $p(\pi_K | \pi_{K-1} \dots \pi_1\})$.  A detailed review of transformers and self-attention is presented in the Appendix \ref{app:transformer}. 
While Decision Transformer \cite{chen2021decision, janner2021offline, zheng2022online} approaches have shown to be effective in certain contexts, surpassing traditional policy gradient (PG) methods by leveraging large amounts of ground truth data, our scenario is distinct in its lack of such ground truth sequences for polar code construction. This absence necessitates a reinforcement learning framework that learns the policy through reward-based feedback.

To streamline the reward generation process in our RL framework, we leverage the open-source C++ AFF3CT toolbox ~\cite{Cassagne2019a} to estimate the BLER rewards efficiently. To further enhance efficiency, we incorporate lookup tables that store and retrieve offline rewards based on previous interactions. A crucial hyperparameter in our framework is the training SNR, specifically chosen to yield a BLER of approximately 0.01 when using the 5G-NR sequence.



%% file: related_works.tex
\section{Related Work}\label{sec:related}
In this section, we review the AI-based approaches for constructing polar codes. In~\cite{ebada2019deep}, a deep neural network (DNN) was used to successfully construct a polar code for a target rate by learning a soft version of the binary mask corresponding to information and frozen positions. Despite its success, this approach lacks the desired nested structure and requires a differentiable decoder.
In another line of work, nested optimization techniques using reinforcement learning (RL) based approaches were proposed in~\cite{elkelesh2019decoder,huang2019ai,huang2019reinforcement,liao2021construction}. RL-based approaches have the unique advantage of not needing a differentiable decoder, and moreover, the nested structure of polar codes is naturally integrated into the problem formulation using RL. All of these learning based approaches provide superior performance compared to the universal sequence by 3GPP and also Gaussian Approximation (GA). It is interesting to note that all these works are limited to the design of polar codes for the Additive White Gaussian Noise (AWGN) channel and do not consider other channels, such as Rayleigh fading, where closed-form approximations such as GA do not exist.  

Our problem formulation is closest to the study by Li et al. ~\cite{li2021learning}, where a variant of the RL-based approach is pursued by transforming it into a policy gradient optimization. An attention mechanism ~\cite{bahdanau2014neural} is used to compute how well the existing information set is aligning with the next best information position, using permutation invariant set representations for ease of learning. In contrast, we parameterize the policy network by a transformer, which can directly predict the next information bit in the nested sequence. Additionally, we highlight that preserving the relative order of reliability by using positional encoding (PE) improves performance. Finally, we consider non-AWGN channels such as Rayleigh fading to and evaluate the benefit of transformers on harder modeling tasks.

%% file: results.tex
\section{Results}\label{sec:results}
In this section, we evaluate the performance of our model for polar code construction of length 64 under the CA-SCL decoder with list size 8. We consider a CRC polynomial 0x3 of length 4. We use Binary Phase Shift Keying (BPSK) modulation and study both AWGN and Rayleigh fading channels.

\textbf{Baselines.} Our benchmarks include the 5G-NR universal reliability sequence described in the 3GPP standards \cite{3gpp2018multiplexing}, and the DE/GA method ~\cite{trifonov2012efficient}. We note the sub-optimality of DE/GA for non-AWGN channels, where the channel gain is not accounted for in the DE/GA approximation. Additionally, we compare our transformer-based model with the attention-based Set-to-Element model~\cite{li2021learning}. For fair comparison with~\cite{li2019pre}, we follow a similar training methodology, details of which are provided in Appendix~\ref{app:train}.

\subsection{Joint Optimization for a target BLER}
We first consider the case of jointly optimizing all rates $\{ \sfrac{1}{N}, \sfrac{2}{N}, \dots ,\sfrac{N}{N} \}$ for a target BLER of 0.01, which is of practical interest for control channel applications. We place equal importance on all rates by choosing $c_k=1$. 

\begin{figure}[!htb]
    \centering
 	\includegraphics[width=1.0\linewidth]{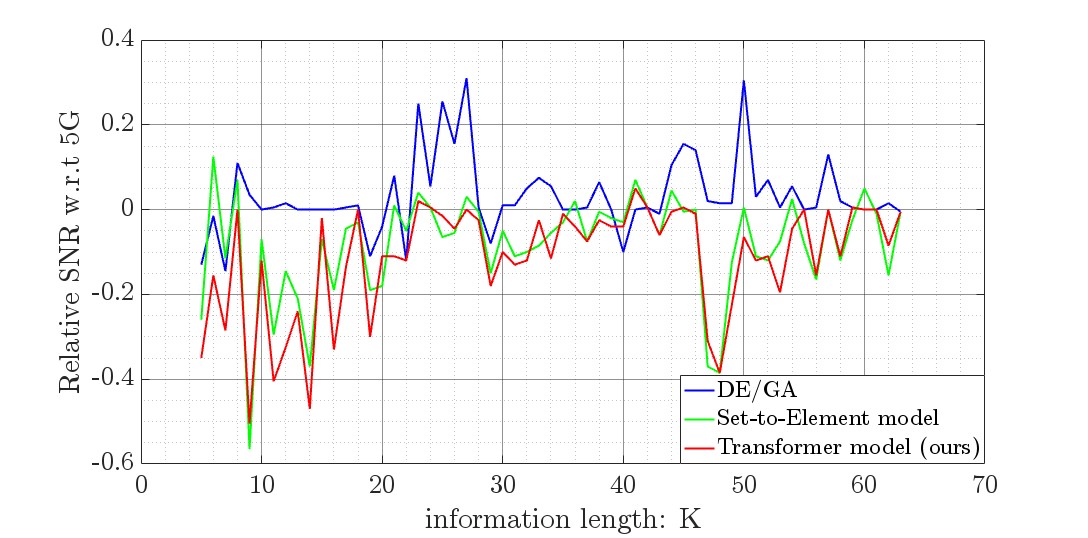}
 	\captionsetup{font=small}
 	\caption{AWGN channel under CA-SCL decoder with list size 8 : Our method shows considerable SNR gain over the 5G-NR sequence and small gains over~\cite{li2021learning}.}
 	\label{fig:awgn_snr}
\end{figure}

In \figref{fig:awgn_snr}, we plot the relative SNR of each scheme with respect to the 5G-NR sequence. We define the relative SNR of a scheme as the SNR required by the scheme minus the SNR required by the 5G-NR sequence. Hence, lower is better. We see in \figref{fig:awgn_snr} that our model provides a gain of 0.05-0.2 dB for various rates with respect to~\cite{li2021learning}. 

\begin{figure}[!htb]
    \centering
 	\includegraphics[width=1.0\linewidth]{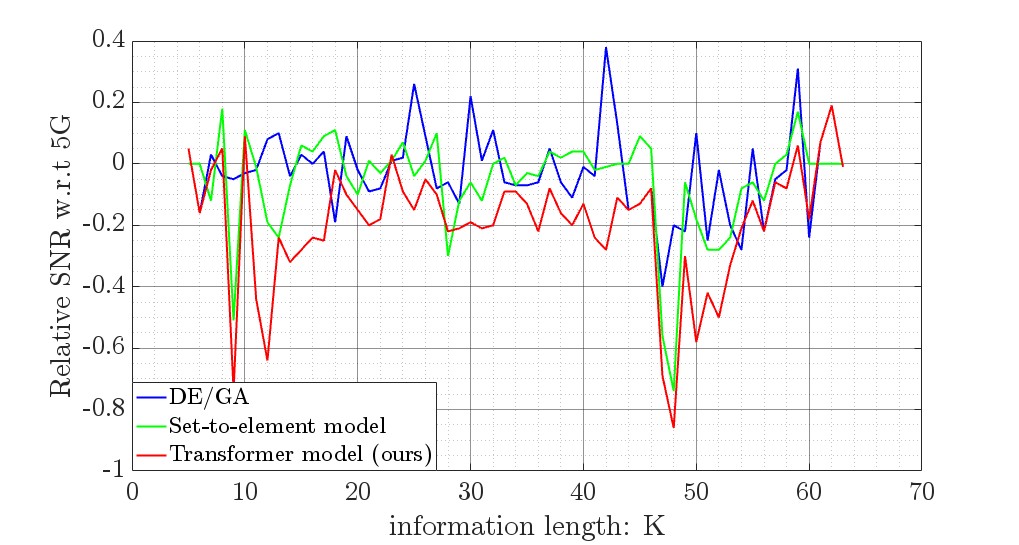}
 	\captionsetup{font=small}
 	\caption{Rayleigh fading channel under CA-SCL decoder with list size 8: Our method shows considerable SNR-gain over the 5G-NR sequence and baselines at target BLER 0.01.}
 	\label{fig:rayleigh_snr}
\end{figure}

Next, we consider the same problem under Rayleigh fading channel with AWGN noise. We see from \figref{fig:rayleigh_snr} that our model gives more noticeable gains of 0.05-0.6 dB compared to~\cite{li2021learning}. We attribute these gains to the superior modeling capabilities of the transformer model. In both cases, our model performs better than 5G-NR and DE/GA methods for most of the rates considered. 

\subsection{Optimization for a target rate }
While all rates can be jointly optimized for a given target BLER, not all rates are of equal importance in practice. It is often desirable to communicate using a fixed {\em set} of rates to keep the complexity low. We consider optimizing for a target rate $\sfrac{K}{N}$ by increasing the weight $c_K$ in the loss term. We use $c_K=10$ in our experiments while $c_k = 1$ for all $k \neq K$.

\begin{figure}[!htb]
    \centering
 	\includegraphics[width=1.0\linewidth]{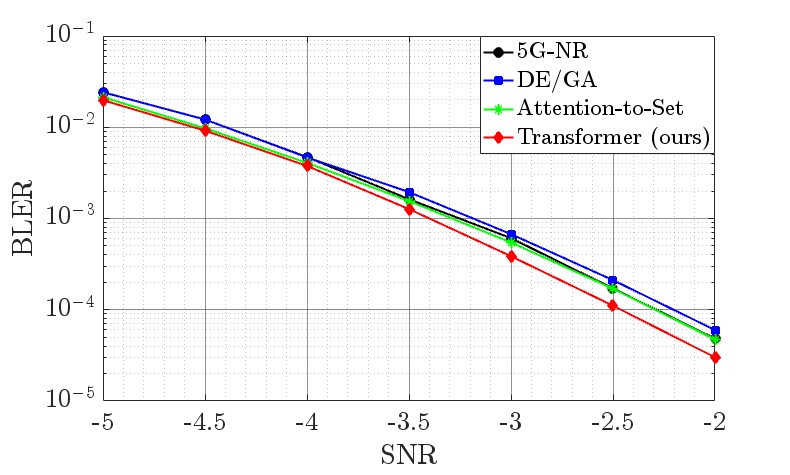}
 	\captionsetup{font=small}
 	\caption{BLER performance in AWGN channel for $(64,16)$ code under CA-SCL decoder with list size 8: For a target rate, our method achieves gains of 0.2-0.3 dB over the 5G-NR sequence and baselines.}
 	\label{fig:awgn_bler}
\end{figure}

In ~\figref{fig:awgn_bler}, BLER performance of rate $\sfrac{1}{4}$ codes (K=16,N=64) in AWGN channel under CA-SCL decoder of list size 8 is plotted. At a BLER of $10^{-4}$, the polar code discovered via our approach is 0.2 dB better than the Attention-to-Set model and 0.3 dB better than the DE/GA model. 

\begin{figure}[!htb]
    \centering
 	\includegraphics[width=1.0\linewidth]{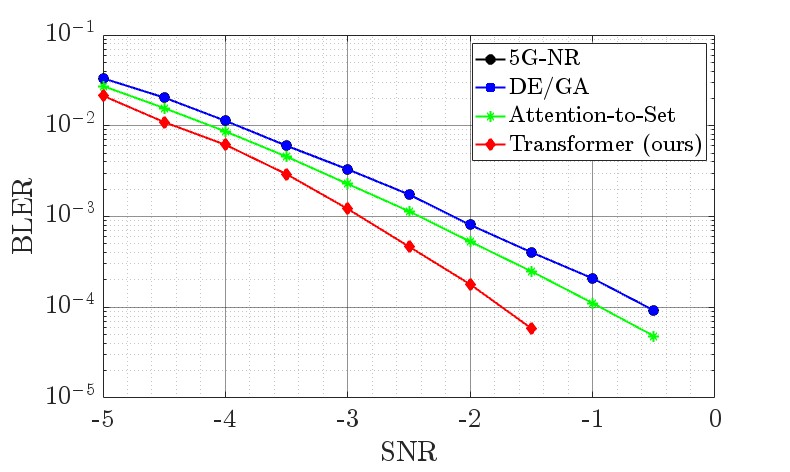}
 	\captionsetup{font=small}
 	\caption{BLER performance in Rayleigh fading channel for $(64,12)$ code under CA-SCL decoder with list size 8: For a target rate, our method achieves gains of 0.8-1.2 dB over the 5G-NR sequence and baselines.}
 	\label{fig:rayleigh_bler}
\end{figure}

In ~\figref{fig:rayleigh_bler}, BLER performance of rate $\sfrac{3}{16}$ codes (K=12,N=64) in Rayleigh fading channel under CA-SCL decoder of list size 8 is plotted.  At a BLER of $10^{-4}$, the polar code discovered via our approach is 0.8 dB better than the Attention-to-Set model and 1.2 dB better than the DE/GA model, demonstrating noticeably higher gains in the case of a more complex modeling task.

\subsection{Effect of permutation invariant representation}
\begin{figure}[!htb]
    \centering
 	\includegraphics[width=1.0\linewidth]{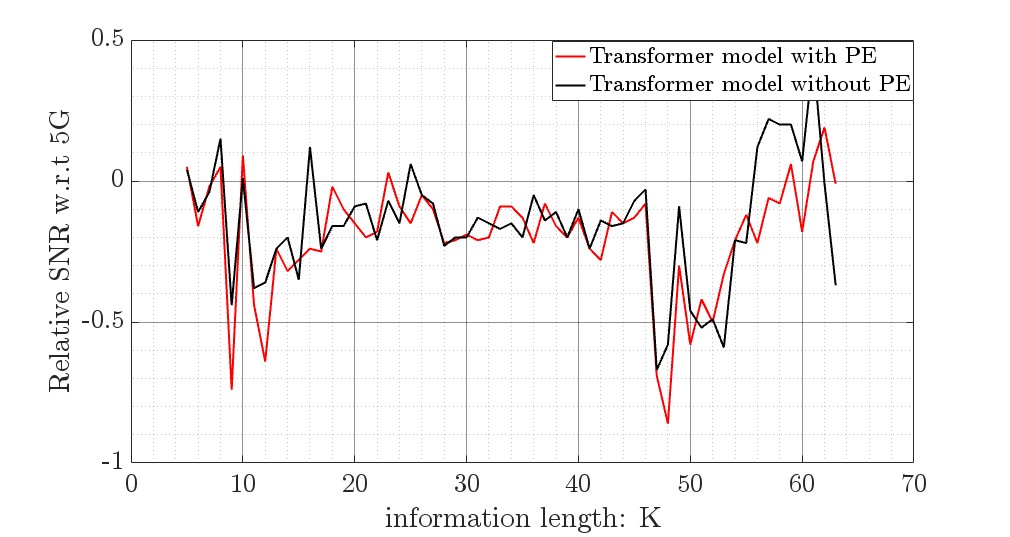}
 	\captionsetup{font=small}
 	\caption{Rayleigh fading channel under CA-SCL decoder with list size 8 at target BLER 0.01: Without PE block, transformer model is experiencing a degradation in performance of up to 0.4 dB.}
 	\label{fig:ablation_rayleigh_snr}
\end{figure}

\begin{figure}[!htb]
    \centering
 	\includegraphics[width=1.0\linewidth]{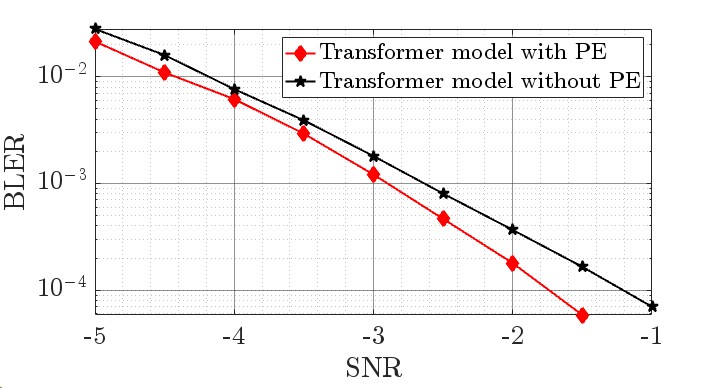}
 	\captionsetup{font=small}
    \caption{Rayleigh fading channel under CA-SCL decoder with list size 8: Without PE block, transformer model is experiencing a degradation in performance of up to 0.35 dB at BLER of $10^{-4}$.}
 	\label{fig:ablation_rayleigh_bler}
\end{figure}

One of the key contributors to the transformer model's success in handling sequential data is the positional encoding (PE) block. To test our hypothesis that preserving the relative positions in the input sequence enhances the quality of learned representations, in contrast to ~\cite{li2021learning}, we conduct an ablation study with the transformer model. \figref{fig:ablation_rayleigh_snr} and \figref{fig:ablation_rayleigh_bler} demonstrate noticeable degradation in the performance of the transformer model when PE is disabled, highlighting the advantage of representations learned by our model over the permutation invariant representations learned in~\cite{li2019pre}. 

%% file: conclusion.tex
\section{Conclusion and Remarks}
In this work, we investigated the problem of nested polar code construction using sequence modeling framework. Specifically, we use a self-attention based transformer to iteratively construct the reliability sequence, one information position at a time. Positional encoding preserves the relative positions of elements in the input sequence and the self-attention mechanism captures the relevance of each input element to other elements in the input sequence, leading to a better estimate of the next element in the sequence. Through empirical results, we demonstrated small gains in terms of BLER and required SNR for AWGN channel and noticeable gains in presence of Rayleigh fading, compared to existing baselines.

There are several interesting future directions for this work. The first is extending the results to larger block lengths. The current bottleneck in this formulation is the reward generation, which is too slow for lengths beyond 64 under CA-SCL decoding. Similarly, the same framework can be employed for designing PAC codes, given the availability of efficient decoder for fast reward generation. Additionally, this approach for code design can be combined with a neural decoder~\cite{hebbar2023crisp, choukroun2022error, aharoni2023data} to optimize the end-to-end code design, while still taking advantage of the inductive bias of polar-like structure that allows for scaling to larger block lengths. 

%% file: bibilography.tex
\vspace{-.8em} 
 \medskip
 \small
 \bibliographystyle{IEEEtran}
 \bibliography{bibilography}

\clearpage
\normalsize

%% file: appendix.tex
\begin{appendices}
\section{Transformer model}\label{app:transformer}
Transformer model is a neural network architecture introduced in~\cite{vaswani2017attention}, geared towards analyzing sequential data. This architecture has since revolutionized many fields fields of AI such as Natural Language Processing (NLP), Computer Vision (CV), time series analysis etc. Two key building blocks of transformer models are the self-attention and positional encoding modules.

Self-attention is a mechanism that allows each element in a sequence to consider and encode information from any other element, thereby capturing the contextual relationships within the sequence. Mathematically, the attention model is defined by the three matrices weight matrices Query ($W_Q$), Key ($W_K$), and Value($W_V$) which are learned during training. To compute these matrices for input embedding $e_i$ as 
\begin{align*}
    Q_i &= W_Q * e_i \\
    K_i &= K_Q * e_i \\
    V_i &= V_Q * e_i
\end{align*}

Once the $W,K,V$ matrices are computed, an attention score can be computed between input $i$ and input $j$ as 
\begin{align*}
    a_{ij} = softmax\left( \frac{Q_j.{K_i}^{T}}{\sqrt{d_K}} \right).
\end{align*}
 to measure the relevance of input $j$ with respect to $i$, where $d_K$ is the size of the embedding.

Additionally, since the transformer's self-attention mechanism treats input tokens independently and doesn't consider their order, positional encoding is added to the input embeddings to provide this crucial sequential context. For instance, positional encoding using sine function can be defined as
\begin{align*}
    PE(pos,2i) = sin \left( \frac{pos}{10000^{\frac{2i}{d_K}}}  \right)
\end{align*}
where $i$ is the position of the element in the sequence and $d_K$ is embedding dimension.

Transformers utilize self-attention and positional encoding in their architecture through multiple attention heads(modules), allowing the model to focus on different parts of the input sequence simultaneously. This parallel processing capability enables Transformers to learn complex dependencies and patterns in data, making them highly effective for tasks such as language translation. Unlike previous sequence models that processed inputs in a linear order (like RNNs and LSTMs), Transformers operate on entire sequences at once. This design eliminates the need for recurrence, leading to significant gains in training efficiency and scalability.

\section{Architecture and Training.}\label{app:train}
As mentioned previously, we desire the neural network to analyze and process the input data \textit{i.e,} the set of sub-channels and their properties, to make decisions \textit{i.e,} selecting the most reliable sub-channels from the available set of sub-channels. Hence, we propose using an encoder only transformer model~\cite{devlin2018bert}. We consider a multi-headed attention with 4 heads, with a 2 layer feedforward network of hidden dimension 256. We employ a learning rate $\in [0.0002, 0.001]$ and train for 20000 epochs with a batch size of 100 samples and decay factor of $\beta = 0.99$ for the moving average of rewards for policy gradient.

\end{appendices}